\documentclass[pdftex,twocolumn,epjc3]{svjour3}          % twocolumn
\RequirePackage[T1]{fontenc}
\smartqed  % flush right qed marks, e.g. at end of proof
\RequirePackage{graphicx}
\RequirePackage{mathptmx}      % use Times fonts if available on your TeX system
\RequirePackage{flushend}
\RequirePackage[numbers,sort&compress]{natbib}
\RequirePackage[colorlinks,citecolor=blue,urlcolor=blue,linkcolor=blue]{hyperref}
\usepackage{graphics}
\usepackage{hyperref}
\usepackage{amsmath}
\usepackage{dirtytalk}
\usepackage{xcolor}
\usepackage{multirow}
\usepackage[normalem]{ulem}
\usepackage{breakurl}
\journalname{Eur. Phys. J. A}

\begin{document}

\title{Isoscaling in central Sn+Sn collisions at 270 MeV/u}

\author{
J.W.~Lee\thanksref{e1,addr1}\and
M.B.~Tsang\thanksref{e2,addr2,addr3}\and
C.Y.~Tsang\thanksref{addr2,addr3}\and
R.~Wang\thanksref{e3,addr2}\and
J.~Barney\thanksref{addr2,addr3}\and
J.~Estee\thanksref{addr2,addr3}\and
T.~Isobe\thanksref{e5,addr4}\and
M.~Kaneko\thanksref{addr4,addr5}\and
M.~Kurata-Nishimura\thanksref{addr4}\and
W.G.~Lynch\thanksref{e4,addr2,addr3}\and
T.~Murakami\thanksref{e6,addr4,addr5}\and
A.~Ono\thanksref{addr6}\and
S.R.~Souza\thanksref{addr7,addr8}\and
D.S.~Ahn\thanksref{addr4}\and
L.~Atar\thanksref{addr9,addr10}\and
T.~Aumann\thanksref{addr9,addr10}\and
H.~Baba\thanksref{addr4}\and
K.~Boretzky\thanksref{addr10}\and
J.~Brzychczyk\thanksref{addr11}\and
G.~Cerizza\thanksref{addr2}\and
N.~Chiga\thanksref{addr4}\and
N.~Fukuda\thanksref{addr4}\and
I.~Gasparic\thanksref{addr12,addr4,addr9}\and
B.~Hong\thanksref{addr1}\and
A.~Horvat\thanksref{addr9,addr10}\and
K.~Ieki\thanksref{addr13}\and
N.~Ikeno\thanksref{addr14}\and
N.~Inabe\thanksref{addr4}\and
G.~Jhang\thanksref{addr2}\and
Y.J.~Kim\thanksref{addr15}\and
T.~Kobayashi\thanksref{addr6}\and
Y.~Kondo\thanksref{addr16}\and
P.~Lasko\thanksref{addr17}\and
H.S.~Lee\thanksref{addr15}\and
Y.~Leifels\thanksref{addr10}\and
J.~\L{}ukasik\thanksref{addr17}\and
J.~Manfredi\thanksref{addr2,addr3}\and
A.B.~McIntosh\thanksref{addr18}\and
P.~Morfouace\thanksref{addr2}\and
T.~Nakamura\thanksref{addr16}\and
N.~Nakatsuka\thanksref{addr4,addr5}\and
S.~Nishimura\thanksref{addr4}\and
H.~Otsu\thanksref{addr4}\and
P.~Paw\l{}owski\thanksref{addr17}\and
K.~Pelczar\thanksref{addr11}\and
D.~Rossi\thanksref{addr9}\and
H.~Sakurai\thanksref{addr4}\and
C.~Santamaria\thanksref{addr2}\and
H.~Sato\thanksref{addr4}\and
H.~Scheit\thanksref{addr9}\and
R.~Shane\thanksref{addr2}\and
Y.~Shimizu\thanksref{addr4}\and
H.~Simon\thanksref{addr10}\and
A.~Snoch\thanksref{addr19}\and
A.~Sochocka\thanksref{addr11}\and
T.~Sumikama\thanksref{addr4}\and
H.~Suzuki\thanksref{addr4}\and
D.~Suzuki\thanksref{addr4}\and
H.~Takeda\thanksref{addr4}\and
S.~Tangwancharoen\thanksref{addr2}\and
Y.~Togano\thanksref{addr13}\and
Z.G.~Xiao\thanksref{addr20}\and
S.J.~Yennello\thanksref{addr18,addr21}\and
Y.~Zhang\thanksref{addr20}
(the S$\pi$RIT collaboration)
}
%\thankstext[$\star$]{t1}{Thanks to the title}
\thankstext{e1}{e-mail: \url{ejungwoo@korea.ac.kr}}
\thankstext{e2}{e-mail: \url{tsang@frib.msu.edu}}
\thankstext{e3}{e-mail: \url{rensheng0611@outlook.com}}
\thankstext{e4}{e-mail: \url{lynch@frib.msu.edu}}
\thankstext{e5}{e-mail: \url{isobe@riken.jp}}
\thankstext{e6}{e-mail: \url{murakami.tetsuya.3e@kyoto-u.jp}}

\institute{Department of Physics, Korea University, Seoul 02841, Republic of Korea\label{addr1}\and
Facility for Rare Isotope Beams, Michigan State University, East Lansing, Michigan 48824, USA\label{addr2}\and
Department of Physics and Astronomy, Michigan State University, East Lansing, Michigan 48824, USA\label{addr3}\and
RIKEN Nishina Center, Hirosawa 2-1, Wako, Saitama 351-0198, Japan\label{addr4}\and
Department of Physics, Kyoto University, Kita-shirakawa, Kyoto 606-8502, Japan\label{addr5}\and
Department of Physics, Tohoku University, Sendai 980-8578, Japan\label{addr6}\and
Instituto de F\'isica, Universidade Federal do Rio de Janeiro, Centro de Tecnologia, Bloco A, 21941-909 Rio de Janeiro, Rio de Janeiro, Brazil\label{addr7}\and
Departamento de Engenharia Nuclear, Universidade Federal de Minas Gerais UFMG, Av. Presidente Antônio Carlos, 6.627, 31270-901 Belo Horizonte, Minas Gerais, Brazil\label{addr8}\and
Institut f\"ur Kernphysik, Technische Universit\"at Darmstadt, D-64289 Darmstadt, Germany\label{addr9}\and
GSI Helmholtzzentrum f\"ur Schwerionenforschung, Planckstrasse 1, 64291 Darmstadt, Germany\label{addr10}\and
Faculty of Physics, Astronomy and Applied Computer Science, Jagiellonian University, Krak\'ow, Poland\label{addr11}\and
Division of Experimental Physics, Rudjer Boskovic Institute, Zagreb, Croatia\label{addr12}\and
Department of Physics, Rikkyo University, Nishi-Ikebukuro 3-34-1, Tokyo 171-8501, Japan\label{addr13}\and
Department of Life and Environmental Agricultural Sciences, Tottori University, Tottori 680-8551, Japan\label{addr14}\and
Rare Isotope Science Project, Institute for Basic Science, Daejeon 34047, Republic of Korea\label{addr15}\and
Department of Physics, Tokyo Institute of Technology, Tokyo 152-8551, Japan\label{addr16}\and
Institute of Nuclear Physics PAN, ul. Radzikowskiego 152, 31-342 Krak\'ow, Poland\label{addr17}\and
Cyclotron Institute, Texas A\&M University, College Station, Texas 77843, USA\label{addr18}\and
Nikhef National Institute for Subatomic Physics, Amsterdam, Netherlands\label{addr19}\and
Department of Physics, Tsinghua University, Beijing 100084, PR China\label{addr20}\and
Department of Chemistry, Texas A\&M University, College Station, Texas 77843, USA\label{addr21}
}

\date{Received: date / Revised version: date}

\maketitle

\abstract{
Experimental information on fragment emissions is important in understanding the dynamics of nuclear collisions and in the development of transport model simulating heavy-ion collisions. The composition of complex fragments emitted in the heavy-ion collisions can be explained by statistical models, which assume that thermal equilibrium is achieved at collision energies below 100 MeV/u. Our new experimental data together with theoretical analyses for light particles from Sn+Sn collisions at 270 MeV/u, suggest that the hypothesis of thermal equilibrium breaks down for particles emitted with high transfer momentum. To inspect the system's properties in such limit, the scaling features of the yield ratios of particles from two systems, a neutron-rich system of ${}^{132}\mathrm{Sn}+{}^{124}\mathrm{Sn}$ and a nearly symmetric system of ${}^{108}\mathrm{Sn}+{}^{112}\mathrm{Sn}$, are examined in the framework of the statistical multifragmentation model and the antisymmetrized molecular dynamics model. The isoscaling from low energy particles agree with both models. However the observed breakdown of isoscaling for particles with high transverse momentum cannot be explained by the antisymmetrized molecular dynamics model. 
}

\clearpage
\section{Introduction}
During a collision involving heavy-ions at energies well above the Coulomb barrier, nuclear matter is driven through very different configurations.
In the early stages, violent collisions between its constituents take place, causing matter to be heated up.
If these collisions occur above the production thresholds, sub-atomic particles such as pions, for instance, may be produced~\cite{StockR_PR_1986_135_259,SengerPeter_PPNP_2004_53_1,OnoAkira_PPNP_2019_105_139_HIC_dynamics,IkenoNatsumi_PRC_2016_93_044612_AMD_SnSn300,HongJun_PRC_2014_90_024605}.
Many particles are ejected in this pre-equilibrium stage~\cite{XuHM_PRC_1994_50_1659}, carrying away an appreciable amount of energy from the system.
At the same time, the relative collective motion between the colliding nuclei leads to the compression of nuclear matter and densities higher than the saturation density found in the core of normal nuclei such as lead~\cite{IkenoNatsumi_PRC_2016_93_044612_AMD_SnSn300,BorderieB_PPNP_2019_105_82_Phase_Transition_Nuclei}.

Dynamical treatments~\cite{WolterHermann_PPNP_2022_125_103962} have been developed to study such early stages. The density, the isospin configuration, and the temperatures attained by the system are sensitive to the physics input parameters employed by the models~\cite{WolterHermann_PPNP_2022_125_103962,XuHM_PRC_1994_50_1659,LiBaoAn_PRL_1997_78_1644,TanWP_PRC_2001_64_051901,JhangG_SpiRIT_PLB_2021_813_136016_pion_ratio}.
Therefore, important information on the Nuclear Equation of State (EOS) may be provided by these dynamical approaches.
Determining the input parameters of the models using experimental observables is of particular interest, 
as some of them are closely related to the rate of equilibration of the system, providing deeper insight into the dynamics of the collisions.

After pre-equilibrium emission, a freeze-out configuration is reached and many fragments including protons and neutrons are emitted~\cite{XuHM_PRC_1994_50_1659}.
Its behaviors can be described both by dynamical~\cite{OnoAkira_PPNP_2019_105_139_HIC_dynamics,WolterHermann_PPNP_2022_125_103962} and statistical models~\cite{BondorfJP_PR_1995_257_133_SMM,DasCB_PR_2005_406_1,BotvinaAS_EPJA_2006_30_121,BorderieB_PPNP_2019_105_82_Phase_Transition_Nuclei}.
In the former case, it appears as a result of the dynamical path taken by the system.
In the case of statistical models, a freeze-out configuration is assumed.
In both cases, most of the excited fragments predicted by the models would decay before detection and, therefore, a de-excitation treatment~\cite{BotvinaAS_NPA_1987_475_663,TanWP_PRC_2003_68_034609_ISMM} must be applied before comparing theoretical predictions with the experimental data.
Consequently, important vestiges of the freeze out configuration may be blurred by the de-excitation process.

In this context, the nuclear isoscaling phenomenon~\cite{TsangMB_MSU_PRC_2001_64_054615_isoscaling_SMM_EES,TsangMB_MSU_PRL_2001_86_5023_isoscaling_experiment}, first reported in Ref.~\cite{XuHS_MSU_PRL_2000_85_716_isoscaling} for Sn+Sn reactions at 50 MeV/u, is a very useful tool as the ratios of yields from two different reactions (which differ mainly in the isospin composition) is weakly affected by the fragment de-excitation process~\cite{TsangMB_MSU_PRC_2001_64_054615_isoscaling_SMM_EES}, retaining information on the system's configuration right after the violent stages of the reaction.
Particularly, under certain conditions~\cite{TsangMB_MSU_PRC_2001_64_054615_isoscaling_SMM_EES,SouzaSR_PRC_2009_80_044606_isoscaling_symmetryenergy,RamiF_PRL_2000_84_1120_FOPI_Isospin_tracing}, it may be related to the symmetry energy, which makes this observable specially relevant to investigations on the nuclear EOS.
The isoscaling analysis considers the ratio:

\begin{equation}
R_{21}(N,Z)=Y_{2}(N,Z)/Y_{1}(N,Z)\;,
\label{eq:ration21}
\end{equation}

\noindent
where $Y_1(N,Z)$ and $Y_2(N,Z)$ stand for the yields of species $(N,Z)$, observed in two similar reactions ``1'' and ``2'' , where $N$ and $Z$ respectively denote the neutron and proton numbers of the isotope. 
It has been found that this ratio follows a simple scaling law \cite{TsangMB_MSU_PRL_2001_86_5023_isoscaling_experiment,TsangMB_MSU_PRC_2001_64_054615_isoscaling_SMM_EES}

\begin{equation}
R_{21}(N,Z)=C\exp (\alpha N + \beta Z),\label{equation_iss}
\end{equation}

\noindent
where $\alpha$ and $\beta$ are the scaling parameters and C is a normalization constant.
By convention, reaction 2 is chosen to have a larger isospin asymmetry compared with that of reaction 1.

This scaling property can be derived in the framework of the grand-canonical ensemble \cite{TsangMB_MSU_PRC_2001_64_054615_isoscaling_SMM_EES}.
From it,
simple relationships between the neutron and proton chemical potentials, $\mu_n$ and $\mu_p$, respectively, and the scaling parameters are obtained:

\begin{equation}
\Delta\mu_n\equiv \mu_n^{(2)}-\mu_n^{(1)}=\alpha T\;
\label{eq:deltamuN}
\end{equation}

\noindent
and

\begin{equation}
\Delta\mu_p \equiv \mu_p^{(2)}-\mu_p^{(1)}=\beta T\;.
\label{eq:deltamuP}
\end{equation}

\noindent
In these expressions, $T$ symbolizes the temperature at the freeze-out configuration and the superscripts ($i=1,2$) represent reactions ``1'' or ``2''.

Although the formal derivation is based on the grand-canonical ensemble, the isoscaling property is also found in nearly all statistical models \cite{TsangMB_MSU_PRC_2001_64_041603_isoscaling_condition}, including versions of the Statistical Multifragmentation Model (SMM) which employ the micro-canonical and canonical ensembles \cite{SouzaSR_PRC_2009_80_044606_isoscaling_symmetryenergy}. It has also been observed in the Antisymmetrized Molecular Dynamics (AMD) model~\cite{OnoAkira_SpiRIT_PRC_2003_68_051601_isoscaling_AMD}. Calculations based on the molecular dynamics approach \cite{DorsoCO_PRC_2006_73_044601} seem to indicate that the isoscaling may be observed even when the system has not yet attained thermal equilibrium. Furthermore, recent experimental results suggest that the parameters $\alpha$ and $\beta$ are sensitive to the mechanisms responsible for fragment production. These results call for further investigations on the nuclear isoscaling property.

Many experimental studies \cite{GeraciE_LNS_NPA_2004_732_173_isoscaling_experiment,TrautmannW_2006_nuclex_0603027_isoscaling_ALADIN_INDRA,FableQ_Arxiv_2022_2202_13850_CaCa_INDRA,WuenschelS_TAMU_PRC_2009_79_061602_isoscaling_experiment,YoungsM_TAMU_NPA_2017_962_61_isoscaling_experiment} have investigated the isoscaling property in different systems at different bombarding energies.
A close relationship between the parameter $\alpha$ and the symmetry energy coefficient $C_\textrm{sym}$ has been pointed out in Ref.\ \cite{TsangMB_MSU_PRC_2001_64_054615_isoscaling_SMM_EES}.
More specifically, denoting by $Z_{i}$ and $A_{i}$ the atomic and mass numbers of the $i$-{\it th} source, one has:

\begin{equation}
\alpha \approx 4C_\textrm{sym}[({Z_1}/{A_1})^2-({Z_2}/{A_2})^2]/T\;.\label{eq:alphaCSym}
\end{equation}

\noindent
This property has been examined in several theoretical \cite{SouzaSR_PRC_2009_80_044606_isoscaling_symmetryenergy,BotvinaAS_JINR_PRC_2002_65_044610_isoscaling_ion_induce} and experimental \cite{TrautmannW_2006_nuclex_0603027_isoscaling_ALADIN_INDRA,FableQ_Arxiv_2022_2202_13850_CaCa_INDRA,WuenschelS_TAMU_PRC_2009_79_061602_isoscaling_experiment,YoungsM_TAMU_NPA_2017_962_61_isoscaling_experiment} works.
Studies found that the isoscaling parameter $\alpha$ and the $(Z/A)^2$ difference between the two sources are related~\cite{WuenschelS_TAMU_PRC_2009_79_061602_isoscaling_experiment}, and different isoscaling tendencies between the projectile-like fragments and the emitted fragments exist~\cite{YoungsM_TAMU_NPA_2017_962_61_isoscaling_experiment}.
The INDRA-GSI collaboration extended this investigation to relativistic collision energies for carbon induced reactions on Sn isotopes \cite{LeFevreA_PRL_2005_94_162701_INDRA_Alladin}.

In this work, we extend earlier investigations at low incident energies of Sn+Sn reactions to a collision energy (270 MeV/u). We concentrate on central mid-rapidity events, with impact parameters $b <1.5$ fm.
The scaling properties are studied as a function of the transverse momentum.

\section{Experiment}

The S$\pi$RIT experiment was performed at the Radioactive Isotope Beam Factory (RIBF) at RIKEN. 
The primary beams of $^{238}$U and $^{132}$Xe impinged on the Be target to produce secondary beams of $^{132,124}\textrm{Sn}$ and $^{112,108}\textrm{Sn}$ respectively, at 270 MeV/u. The beams bombarded on the isotopically enriched $^{124}\textrm{Sn}$ and $^{112}\textrm{Sn}$ targets with areal density of $608\,\textrm{mg/cm}^2$ and $561\,\textrm{mg/cm}^2$, respectively. Four reactions with different neutron-to-proton ratios, $N/Z$, of the total system were measured:
${}^{132}\mathrm{Sn}+{}^{124}\mathrm{Sn}$ $(N/Z = 1.56)$,
${}^{108}\mathrm{Sn}+{}^{112}\mathrm{Sn}$ $(N/Z = 1.2)$,
${}^{112}\mathrm{Sn}+{}^{124}\mathrm{Sn}$ $(N/Z = 1.36)$, and
${}^{124}\mathrm{Sn}+{}^{112}\mathrm{Sn}$ $(N/Z = 1.36)$.
In this work, we focus mainly on the most $(N/Z = 1.56)$ and least $(N/Z = 1.2)$ neutron rich systems.

Charged particles produced in the reactions were detected with the SAMURAI Pion Reconstruction and Ion-Tracker Time Projection Chamber (S$\pi$RIT-TPC)~\cite{ShaneR_SpiRIT_NIMA_2015_784_513_spirittpc,TangwancharoenS_SpiRIT_NIMA_2017_853_44_TPCGatingGrid,BarneyJ_SpiRIT_RSI_2021_92_063302_spirittpc} installed inside the SAMRURAI dipole magnet~\cite{OtsuH_NIMB_2016_376_175_SAMURAIMagnet} with a magnetic field of 0.5 T. 
The effective volume of the TPC is 1344 mm $\times$ 864 mm $\times$ 530 mm, and the target was placed at 15 mm upstream of the entrance window, resulting in an angular coverage of $\theta<80^\circ$ with respect to the beam axis in the laboratory frame.
Description of the associated trigger arrays located on the side and downstream of the TPC used to select central events can be found in Refs.~\cite{LaskoP_NIMA_2017_856_92_KATANA,KanekoM_NIMA_2022_1039_167010_KyotoArray}. The Generic Electronics for TPCs (GET) was employed to read out the track signals. The analysis software S$\pi$RITROOT~\cite{JhangGenie_SpiRIT_JKPS_2016_69_144_SpiRITROOT,LeeJW_SpiRIT_NIMA_2020_965_163840_SpiRITROOT} was developed for the track reconstruction of the charged particles. 
Detailed performance of the TPC and GET electronics as well as the  software analysis codes have been published in ~\cite{BarneyJ_SpiRIT_RSI_2021_92_063302_spirittpc,JhangGenie_SpiRIT_JKPS_2016_69_144_SpiRITROOT, IsobeT_SpiRIT_NIMA_2018_899_43_GET_electronics, LeeJW_SpiRIT_NIMA_2020_965_163840_SpiRITROOT}. Other technical issues including space charge correction and extending the dynamic range of the TPC electronics can be found in~\cite{TsangCY_SpiRIT_NIMA_2020_959_163477_space_charge, EsteeJ_SpiRIT_NIMA_2019_944_162509_tpcdynamicrange}.

\begin{figure}
\centering
\includegraphics[width=1.0\hsize]{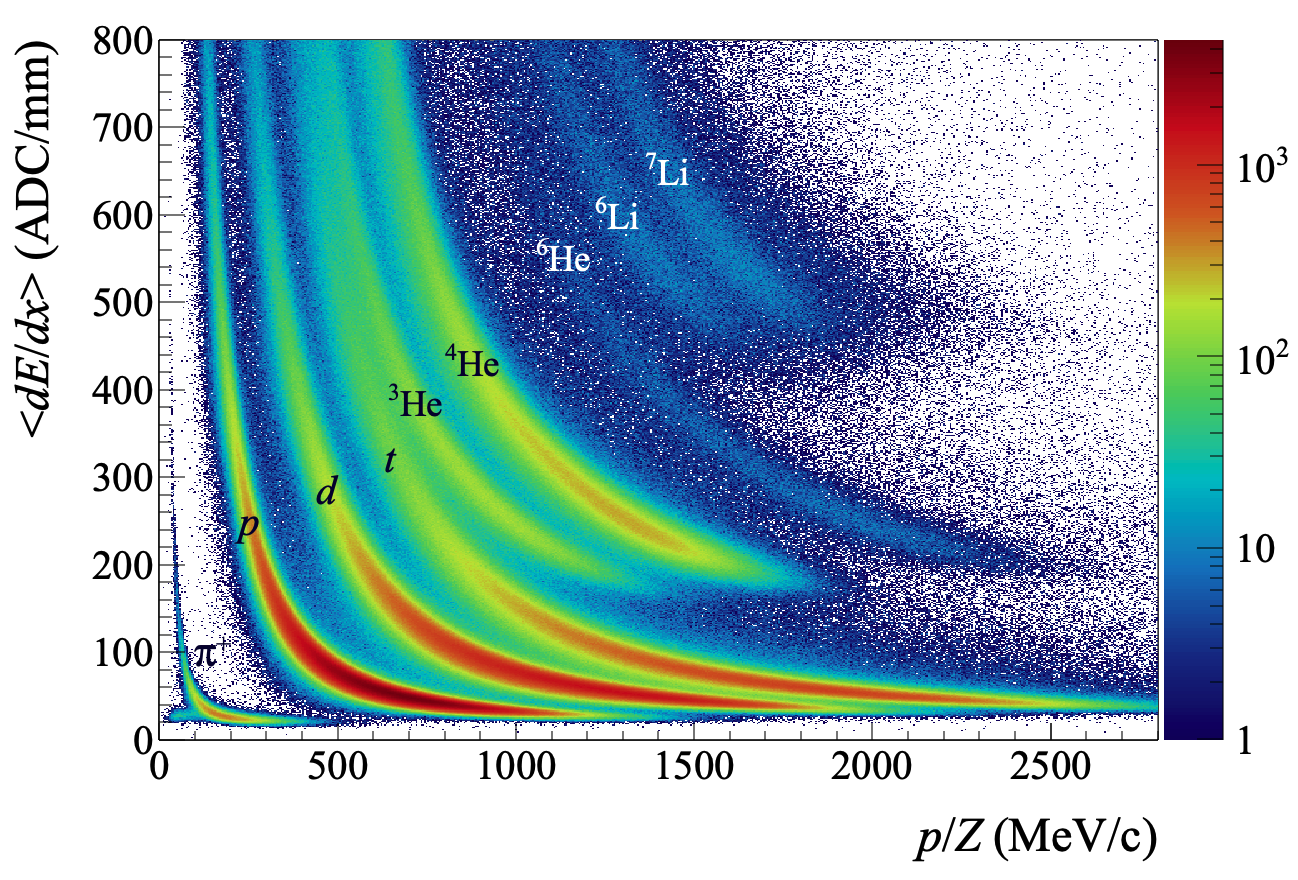}
\caption{Particle identification plot of energy loss, $\left<dE/dx\right>$ vs. magnetic rigidity, $p/Z$. The particles emitted from the $^{132}\textrm{Sn}+^{124}\textrm{Sn}$ collisions at 270 MeV/u are detected with the S$\pi$RIT-TPC.}\label{figure_pid}
\end{figure}

The cuboid-shaped S$\pi$RIT TPC lacks the azimuthal symmetry. In this work, we mainly analyze the data at the azimuthal angles, $\phi$, $-30^\circ<\phi<20^\circ$ and $160^\circ<\phi<210^\circ$ where the tracks are longest and the track quality is generally much better. 
The TPC efficiencies arising from the detector performance are determined by the track embedding method~\cite{AndersonM_NIMA_2003_499_659_STARTPC} using events generated from Monte Carlo simulations and GEANT-4. The reconstructed rigidity momentum/charge, $p/Z$ and the mean energy loss per unit length $\left<dE/dx\right>$ were provided for each track. The resolution of the reconstructed momentum and $\left<dE/dx\right>$ for single tracks is 1.6\% and 4.6 \%, respectively~\cite{EsteeJustinBrian_phdthesis_MichiganStateUniversity_2020}. The particle identification (PID) is obtained using the correlation plot of magnetic rigidity, $p/Z$, and energy loss, $\left<dE/dx\right>$, of the detected particles, as shown in Fig.~\ref{figure_pid}. Isotopes of $p$, $d$, $t$, ${}^{3}\mathrm{He}$, $^{4}$He, ${}^{6}\mathrm{He}$, ${}^{6}\mathrm{Li}$, and ${}^{7}\mathrm{Li}$ can be clearly identified. However, below $p/Z <$ 600 MeV/$c$, the $t$ and ${}^{3}\mathrm{He}$ PID lines merge. In this study, we mainly focus on particles in the mid-rapidity range $y_{0} =$ 0 - 0.4 with $y_{0} = y/y_{NN}^{c.m.} - 1$ where $y$ is the rapidity of the particle and $y_{NN}^{c.m.}$ is center-of-mass rapidity of the nucleon-nucleon system. 
Since the mid-rapidity gate eliminates tritons below $p/Z <$ 1000 MeV/$c$, contamination from ${}^{3}\mathrm{He}$ in triton spectra is minimal. On the other hand, contamination from triton in the $^{3}$He spectra has to be determined.

Some physics results regarding the symmetry energy constraints from the S$\pi$RIT experiments have been published. Charged pion multiplicities and ratios were published in Ref.~\cite{JhangG_SpiRIT_PLB_2021_813_136016_pion_ratio} and Ref.~\cite{EsteeJ_SpiRIT_PRL_2021_126_162701_pion_ratio}. Ref.~\cite{KanekoM_SpiRIT_PLB_2021_822_136681_Z1particles_AMD} focused on $Z$=1 particles and comparisons of the rapidity distributions with the AMD models\cite{OnoAkira_PTP_1992_87_1185_AMD,OnoAkira_PPNP_2019_105_139_HIC_dynamics}. This paper provides a more comprehensive study than Ref.~\cite{KanekoM_SpiRIT_PLB_2021_822_136681_Z1particles_AMD} and focuses on the transverse momentum spectra and yield ratios obtained from light charged fragments, specifically $p$, $d$, $t$, $^{3}$He and $^{4}$He.   

To select central collisions, we assume the impact parameters increase monotonically with the charged particle multiplicity~\cite{CavataC_PRC_1990_42_1760_reduced_impact_parameter, BarneyJonathanElijah_phdthesis_MichiganStateUniversity_2019}. Very central collision events ($b <$ 1.5 fm) similar to those analyzed in Ref.~\cite{KanekoM_SpiRIT_PLB_2021_822_136681_Z1particles_AMD} are chosen for this work. 
The difference between this work and Ref.~\cite{KanekoM_SpiRIT_PLB_2021_822_136681_Z1particles_AMD} for the rapidity spectra in $y_{0} =$ 0 - 0.4 (Fig.~\ref{figure_dmdy}) is within 3~\%. The charge multiplicity selection in both works is different. Indeed, events with multiplicities equal or more than 57 and 56 charged particles, for $^{132}$Sn+$^{124}$Sn and $^{108}$Sn+$^{112}$Sn respectively, are chosen in this work, while 56 and 55 are used in Ref.~\cite{KanekoM_SpiRIT_PLB_2021_822_136681_Z1particles_AMD}. In both cases, the impact parameter gates of ($b <$ 1.5 fm) is chosen. In this work, $b_\mathrm{max}$ is defined to be the experimental $b_\mathrm{max}$ of 7.52 fm in ${}^{132}\mathrm{Sn}+{}^{124}\mathrm{Sn}$ and 7.13 fm in ${}^{108}\mathrm{Sn}+{}^{112}\mathrm{Sn}$ while in Ref.~\cite{KanekoM_SpiRIT_PLB_2021_822_136681_Z1particles_AMD}, $b_\mathrm{max}$=10 fm is defined by the sum of the radii for the projectile and target. These slight differences in the multiplicity gates do not affect the results of both works. The total statistics is increased as the tracks from both the left and right side of the TPC is used in the present work, whereas only the tracks in the right side of the TPC is used in Ref.~\cite{KanekoM_SpiRIT_PLB_2021_822_136681_Z1particles_AMD}. These are the main reasons for the 3~\% discrepancy. In any case, the isocaling ratios are unaffected by the slight discrepancies in the spectra.

Fig.~\ref{figure_pid} shows a typical PID plot of the charged particles obtained in the $^{132}$Sn+$^{124}$Sn collisions. The heavier clusters with A $>$ 5 nuclei such as $^{6}\textrm{He}$, $^{6}\textrm{Li}$, $^{7}\textrm{Li}$ are not included in the current analysis due to lack of statistics. To standardize the PID conditions, the center and width of $\left<dE/dx\right>$ of the PID line for each particle are fitted with the empirical Bethe-Bloch formula with Gaussian widths. Only tracks having PID probability larger than $70\%$ ($50\%$ for $^{3}\textrm{He}$) and $2.2\sigma$ width of $\left<dE/dx\right>$ are selected for the analysis.

%\begin{table*}
%\centering
%\small
%{
%\begin{tabular}{llll}
%\hline
%Cut &This work & Ref.~\cite{KanekoM_SpiRIT_PLB_2021_822_136681_Z1particles_AMD} \\ \hline
%Maximum impact parameter $b_\textrm{max}$ & 7.52 & 11.59 & (${}^{132}\mathrm{Sn}+{}^{124}\mathrm{Sn}$)\\
%                                          & 7.13 & 11.02 & (${}^{108}\mathrm{Sn}+{}^{112}\mathrm{Sn}$)\\
%%Azimuthal angle & $-30^\circ<\phi<20^\circ$ and $160^\circ<\phi<210^\circ$ & $-30^\circ<\phi<20^\circ$ \\
%\hline
%\end{tabular}
%}
%\caption{List of applied cuts used in this work and in %Ref.~\cite{KanekoM_SpiRIT_PLB_2021_822_136681_Z1particles_AMD}.}
%\label{table_ana_cuts}
%\end{table*}

The overall systematic uncertainties are estimated from the variations of track multiplicity, track quality~\cite{LeeJW_SpiRIT_NIMA_2020_965_163840_SpiRITROOT} and PID quality cuts~\cite{TsangChunYuen_phdthesis_MichiganStateUniversity_2022}. 
In the region $p_T/A <$ 400 MeV/c, the statistical errors are smaller than systematic errors which
are about 5\% and 2\% for the absolute yield and yield ratios, respectively. 
Outside of this region the statistic and systematic uncertainties increase with $p_T/A$ to exceed 15\%. The error bars shown in this work include both the systematic and statistical errors.

\section{Particle Spectra}

\begin{figure}
\centering
\includegraphics[width=1.0\hsize]{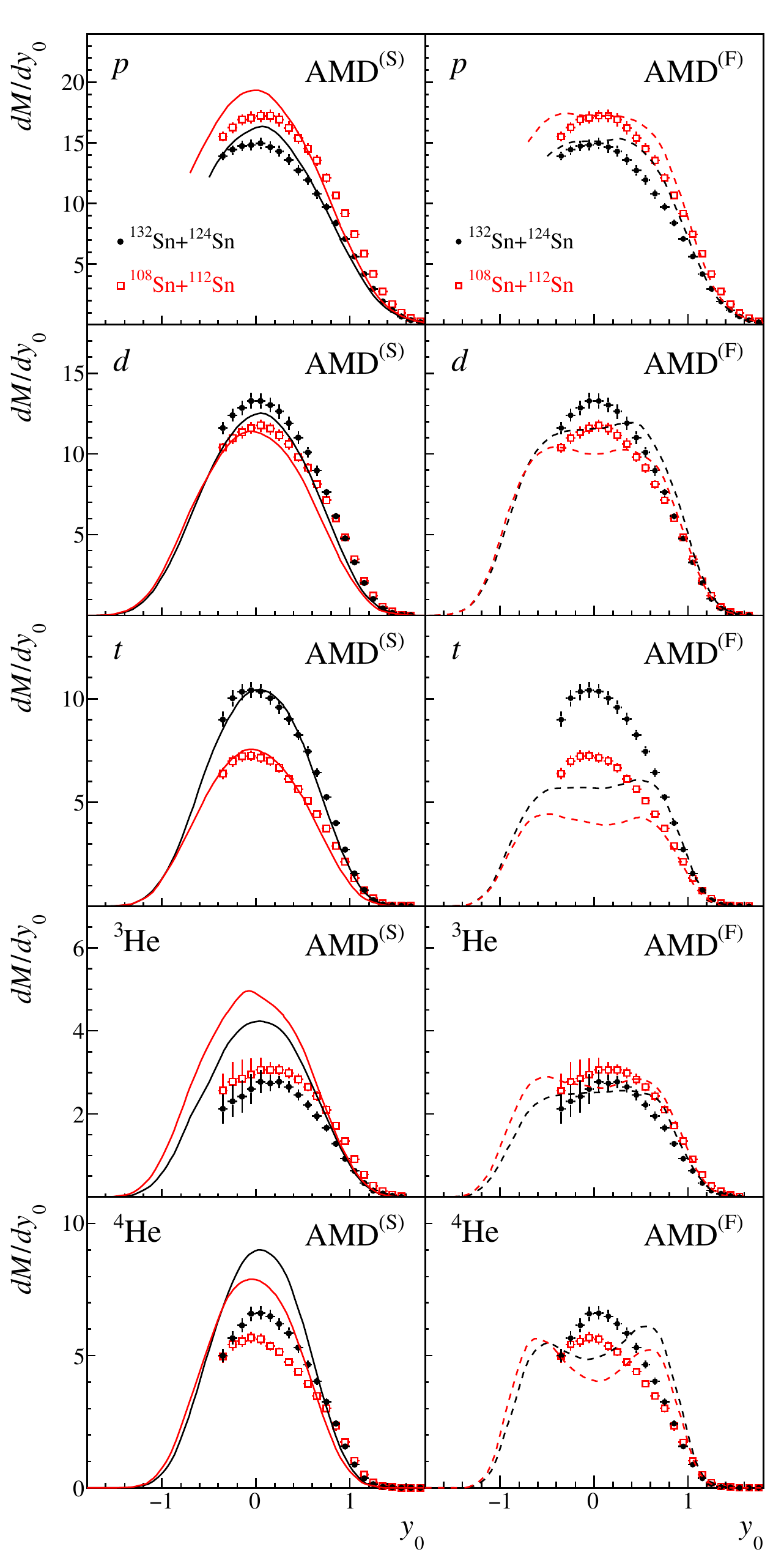}
\caption{Differential multiplicity as a function of rapidity $y_0$ for $p$, $d$, $t$, $^3$He, and $^4$He from top to bottom panels from the collisions of ${}^{132}\mathrm{Sn}+{}^{124}\mathrm{Sn}$ (black solid circles) and ${}^{108}\mathrm{Sn}+{}^{112}\mathrm{Sn}$ (red open squares) reactions. The data points are compared to the AMD$^{(\textrm{S})}$ (solid lines) on the left panels and AMD$^{(\textrm{F})}$ (dotted lines) on the right panels. }\label{figure_dmdy}
\end{figure}
 
\begin{figure}
\centering
\includegraphics[width=1.0\hsize]{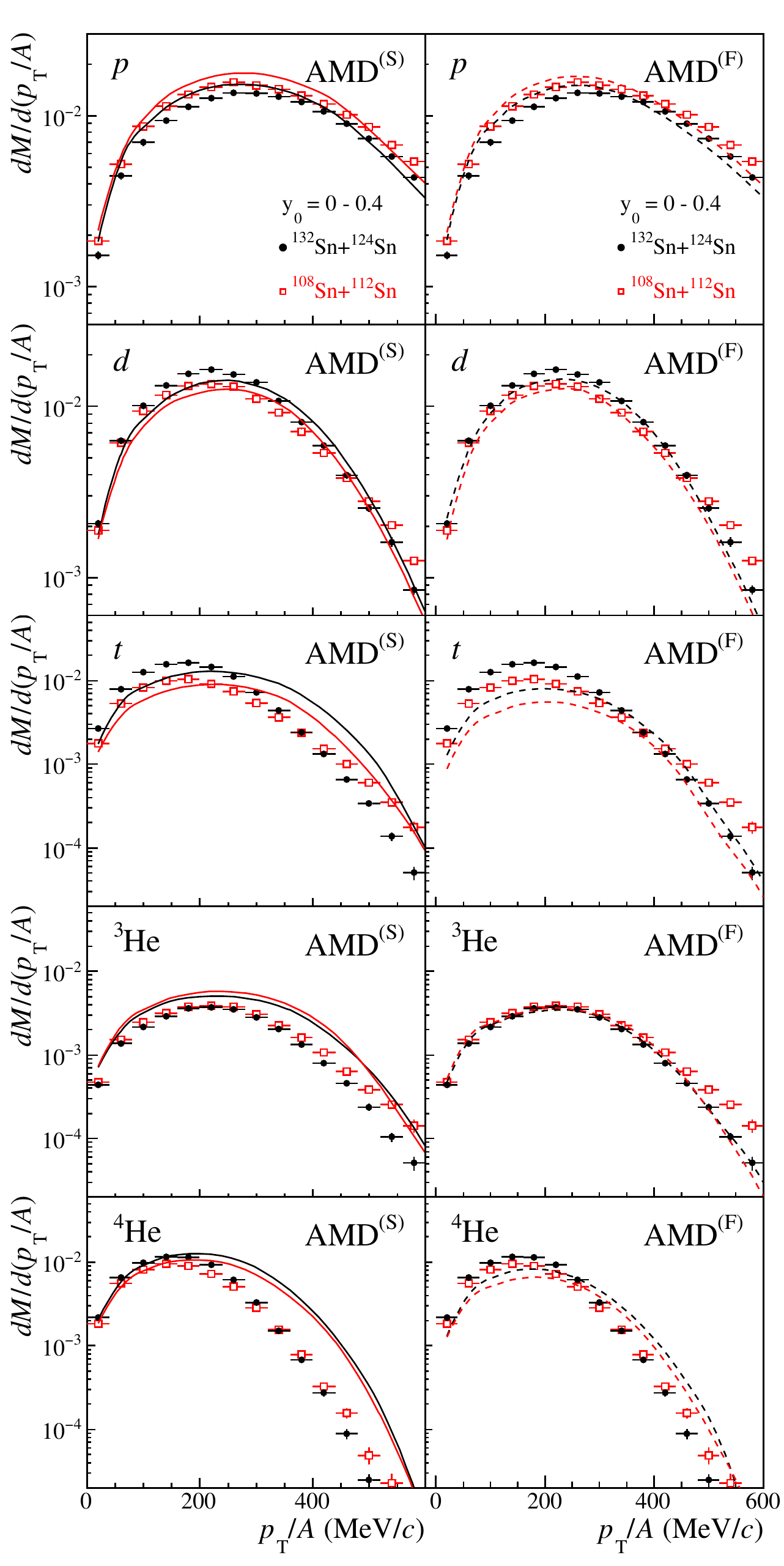}
\caption{Differential multiplicity as a function of $p_T/A$ for $p$, $d$, $t$, $^3$He, and $^4$He from top to bottom panels. Data points are compared to the AMD$^{(\textrm{S})}$ (solid lines) on the left panels and AMD$^{(\textrm{F})}$ (dotted lines) on the right panels for ${}^{132}\mathrm{Sn}+{}^{124}\mathrm{Sn}$ (black) and ${}^{108}\mathrm{Sn}+{}^{112}\mathrm{Sn}$ (red) reactions.}\label{figure_dmdp}
\end{figure}

The experimental rapidity and transverse momentum spectra for $Z$=1 and 2 particles are shown in Fig.~\ref{figure_dmdy} and Fig.~\ref{figure_dmdp}, respectively, for the neutron rich system ${}^{132}\mathrm{Sn}+{}^{124}\mathrm{Sn}$ $(N/Z = 1.56)$ (black circles) and the nearly symmetric system, ${}^{108}\mathrm{Sn}+{}^{112}\mathrm{Sn}$ $(N/Z = 1.20)$ (red squares). The data in both left and right panels in each figure are the same. Comparison of the data to the two different parameter sets of the AMD models, shown by solid (left panels) and dashed (right panels) lines, will be discussed in details later.

In all cases, the rapidity spectra show peaking near or at $y_0 = 0$ suggesting high degree of stopping. The slight asymmetry observed around $y_0 = 0$ for all particles is due to the inefficiencies in the TPC to detect target rapidity particles which are generally low in energy and emitted at backward angles in the laboratory frame. In the case of $^{3}\mathrm{He}$, the PID contamination from tritons is significant and the  spectra peak is located slightly off $y_0 = 0$. For the analysis, we assume the spectra is symmetric at $y_{0}=0$ and only include data in $y_{0}=0$ - $0.4$.

At mid rapidity, except for proton and $^{3}\mathrm{He}$ isotopes, more particles, $d$, $t$ and $^{4}\mathrm{He}$ are produced in the neutron-rich systems. The same observation is also seen in  Fig.~\ref{figure_dmdp} for particles with $p_{T}/A < 280$ MeV/$c$. On the contrary, more high energy particles including the neutron rich tritons, with $p_{T}/A > 400$ MeV/$c$,  are produced from the nearly symmetric ${}^{108}\textrm{Sn}+{}^{112}\textrm{Sn}$ system than the neutron-rich one. This is surprising as one would expect that neutron-rich systems would produce more neutron-rich isotopes at all energies. One explanation could be that high energy particles are produced in a dynamical and non-equilibrium environment and that the hot participant zone is rather neutron deficient. 

\section{Yield Ratios and Isoscaling}

It has been shown that isoscaling occurs only when the two systems have nearly the same temperature~\cite{TsangMB_MSU_PRC_2001_64_041603_isoscaling_condition}. Since the absolute temperature cannot be directly measured, we use the established isotope thermometers based on the double ratio of hydrogen and helium species

\begin{equation}
R_{\textrm{H}-\textrm{He}}=[Y(d )Y(^4\textrm{He})]/[Y(t)Y(^3\textrm{He})]
\label{eq:RH}
\end{equation}

\noindent
to examine the relative temperature increase with particle energy in both reactions: 

\begin{equation}
T_{\textrm{H}-\textrm{He}} \,\,\textrm{(MeV)} = \frac{14.29}{\log\left(1.59 \,R_{\textrm{H}-\textrm{He}}\right)}\label{equation_temperature}\;.
\end{equation}
We note that the isotope temperature can be derived in the grand-canonical ensemble~\cite{AlbergoS_INCA_1985_89_1} and it is found to be independent of $N/Z$ ratio of the source in low energy collisions~\cite{KundeGJ_PLB_1998_416_56_HHeTemperature}.

\begin{figure}
\centering
\includegraphics[width=0.95\hsize]{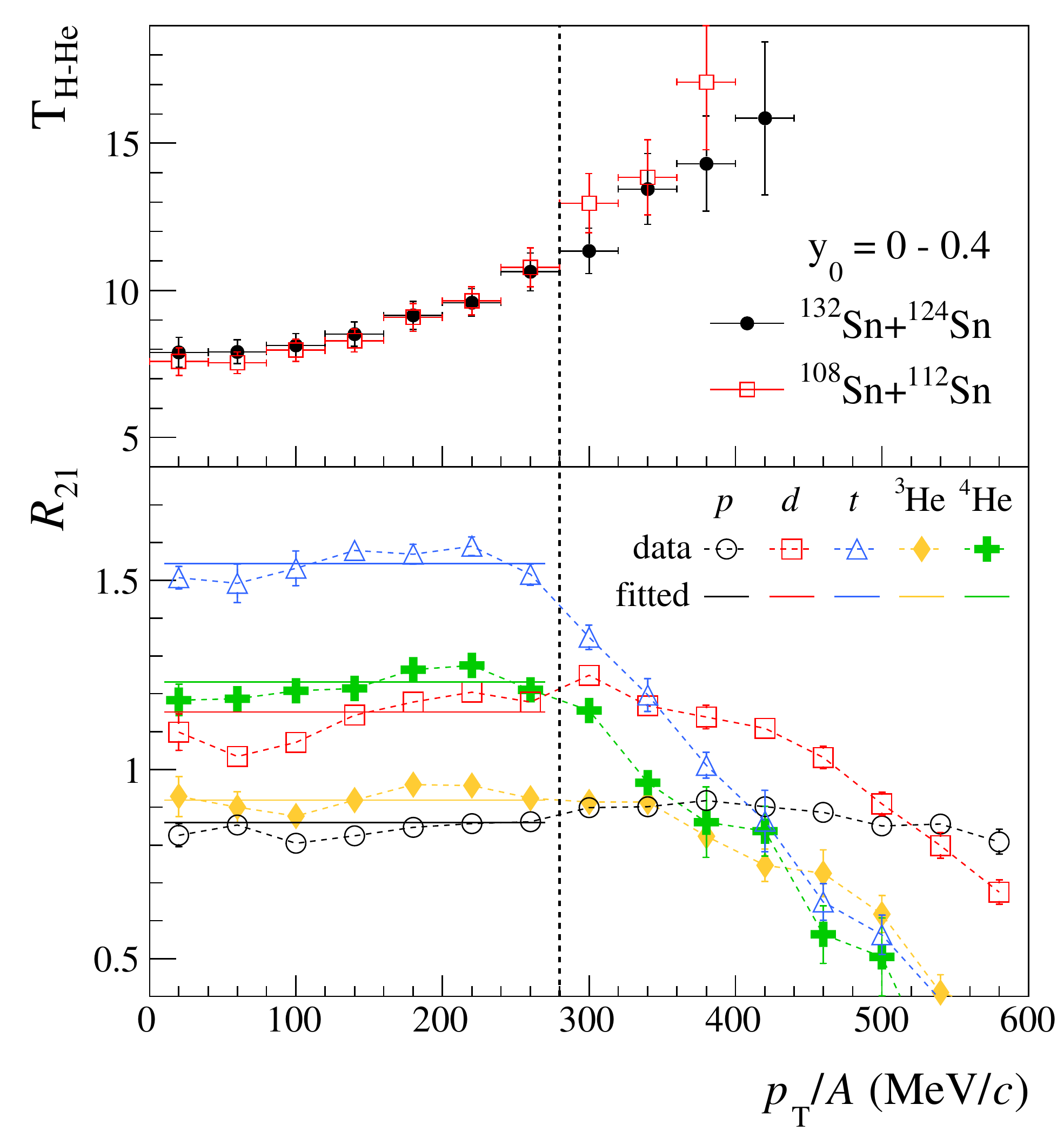}
    \caption{Upper panel: The H-He temperature evaluated using the isotope ratios of Eq.~\ref{eq:RH} \& \ref{equation_temperature}. Lower panel: The isoscaling yield ratio $R_{21}$ as a function of transverse momentum for different particles. The break line at $p_T/A$ = 280 MeV/$c$ show  different trend of isoscaling for particles with low (left side of the line) and high (right side of the line) $p_T/A$. The horizontal solid lines for $p_T/A <$ 280 MeV/$c$ correspond to the average fitted values of $R_{21}$ using Eq.~\ref{equation_iss}. The dashed lines are used to guide the eye highlighting the trends of the data for different particles.}
    \label{figure_r21_temp}
\end{figure}

The upper panel of Fig.~\ref{figure_r21_temp}
displays the H-He temperature evaluated from Eq.~(\ref{equation_temperature}).
The $T_\textrm{H-He}$ for both neutron rich (solid circles) and the near-symmetric (open squares) systems are nearly the same for low energy particles. The temperatures increases slowly from around 8 MeV to 10 MeV with increasing $p_T/A$. Above 280 MeV/$c$, temperatures start to increase  dramatically. Furthermore, the temperatures of the two systems begin to differ. Above 400 MeV/$c$ the temperature of the near-symmetric system is higher than that of the neutron rich system. 
The increasing differences in temperatures with $p_T/A$ indicates that the Sn+Sn collisions studied here (with incident energy of 270 MeV/u) do not form fully equilibrated systems. 
%Such behavior was not observed in previous work~\cite{ChajeckiZ_MSU_Arxiv_2014_1402_5216_isotopicratio} at lower beam energy.

The increase in H-He temperature as a function of the surface velocity of the particles, has been observed in  Refs.~\cite{WangJ_PRC_2005_72_024603,BougaultR_JPG_47_025103}. In previous work, a drop of the temperature is also observed at very high velocity. We do not see the drop. This could be a consequence of the dramatic decrease of the cross-sections for $p_T/A$ above 280 MeV/$c$.    

Next, we focus on the spectral ratio of isotope yields $R_{21}(N,Z)$ of the $Z$=1 and 2 particles, observed in the ${^{132}\textrm{Sn}}+{^{124}\textrm{Sn}}$ and ${^{108}\textrm{Sn}}+{^{112}\textrm{Sn}}$ systems. These are shown in the lower panel of Fig.~\ref{figure_r21_temp} as a function of $p_T/A$. The dotted lines connect data points providing visual guidance to the trends of the data for each particle. The vertical dashed line at $p_T/A = 280$ MeV/$c$ marks the approximate region when the temperatures of the two systems start to differ. 

The isotope ratios plotted to the left of the dashed line show isoscaling characteristics. In each ($N-Z=-1$, 0 and 1) group, $R_{21}(N,Z)$ values are nearly constant as a function of $p_T/A$. Moreover, $^3$He behave like protons ($N-Z=-1$) with ratio value smaller than 1, and $^4$He behave similarly to deuterons ($N-Z=0$) with higher ratio value than protons. The tritons ($N-Z=1$) have the highest ratio values as expected from isoscaling. 

\begin{figure}
\centering
\includegraphics[width=1.0\hsize]{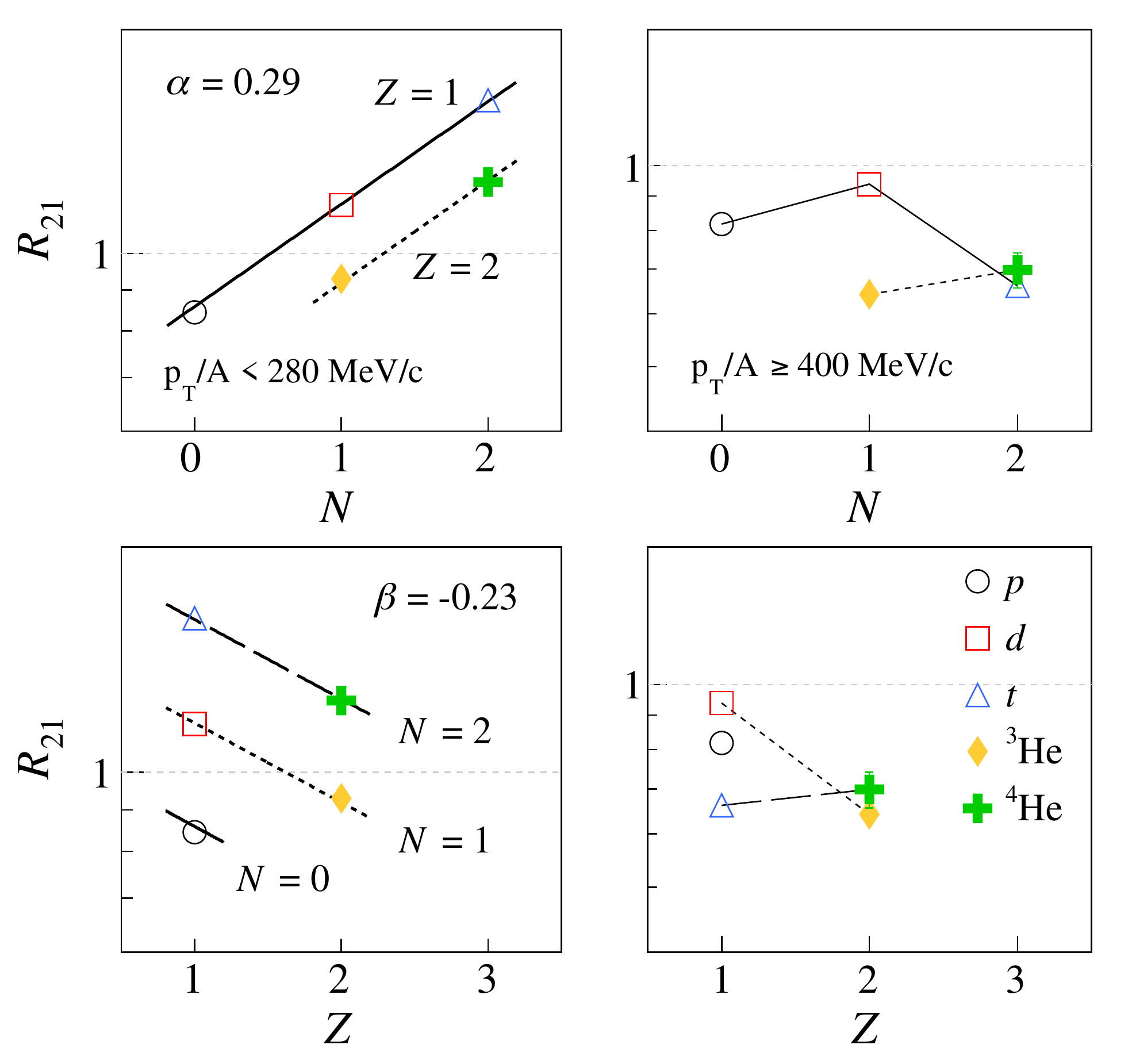}
\caption{
In each panel, $R_{21}$ values of Z=1 and 2 particles are plotted as a function of $N$ and $Z$ in the top and bottom panels, respectively. 
In the left panels, the lines are fits with Eq.~(\ref{equation_iss}) performed for particles with $p_T/A <$ 280 MeV/$c$.  The values of $\alpha$ and $\beta$ noted on the top corners of the panels correspond to the slope of the fitted lines.  In the right panels, the data shown for $pT/A \geq$ 400 MeV/$c$ do not follow Eq.~(\ref{equation_iss}), and the lines joining the data points are mainly used to guide the eyes.
}\label{figure_fit}
\end{figure}

The $R_{21}$ ratios for the five particles with $p_T/A <$ 280 MeV/$c$  are plotted as a function of $N$ and $Z$, in the left panels of Fig~\ref{figure_fit}. The three parameters $\alpha$, $\beta$ and $C$ are simultaneously fitted and the resulting isoscaling fits are shown as lines in the figure. The two fitted lines of the data for the isotopes with $Z$ = 1 and 2 are shown separately in the top left panel, where the slopes of the lines represent the fit parameter $\alpha$ = 0.29. 
The three fitted lines for the isotones with $N$ = 0, 1 and 2 are shown on the bottom left panel as a function of $Z$ and the slopes represent the fit parameter $\beta$ = -0.23. In the absence of Coulomb, protons and neutrons should behave similarly and one would expect $\alpha$ and $\beta$ to have similar values but opposite signs. This has been observed in most of the previous studies of isoscaling. In this case, the magnitude of the $\alpha$ value is larger than the $\beta$ value. The isoscaling ratios obtained from these fits are plotted as solid horizontal lines in the bottom panel of Fig.~\ref{figure_r21_temp} to the left of the vertical dashed line. 

In contrast, except for protons, $R_{21}(N,Z)$ for high energy isotopes shown on the right side of the vertical dashed line decreases with $p_T/A$.  Furthermore, the lines from different isotopes cross over each other. In the case of $N=2$ particles (triton and $^4$He), $R_{21}$ falls off suddenly above $p_T/A =$ 280 MeV/$c$ with the largest drop exhibited by tritons. This reflects the sharper drop in the $t$ and $^4$He particle spectra at the high energy. 
The right panels of Fig.~\ref{figure_fit} show the ratios $R_{21}$ plotted as a function of $N$ and $Z$ in the range of $p_{T}/A \geq 400$ MeV/$c$. $R_{21}$ values for t, ${}^{3}\mathrm{He}$ and $^{4}$He have nearly the same values leading to the breakdown of isoscaling. The lines connecting the isotope (upper right panel) and isotone (lower right panel) data points serve only to guide the eye and provide a contrast trends of the $R_{21}$ values between fragments with low (left panels) and high (right panels) transverse momentum.

It is also interesting to note that $R_{21}$ is less than 1 for all high energy isotopes. For isotopes with $p_{T}/A \approx 500$ MeV/$c$, $R_{21} \approx$ 0.5 for t, $^{3}$He and $^{4}$He, {\it i.e.} 50\% less tritons are produced from the neutron-rich system of ${^{132}\textrm{Sn}}+{^{124}\textrm{Sn}}$ than that from the ${^{108}\textrm{Sn}}+{^{112}\textrm{Sn}}$ system. So far there is no explanation for the surprising result.

\section{Model Comparisons}

The isoscaling ratios $R_{21}(N,Z)$ of the $Z$ = 1 and 2 particles, obtained from the experimental yields measured in the ${^{132}\textrm{Sn}}+{^{124}\textrm{Sn}}$ and ${^{108}\textrm{Sn}}+{^{112}\textrm{Sn}}$ systems, are shown in Fig.~\ref{figure_r21}. In each panel, data are compared to the models; SMM (horizontal lines in the top panel), AMD$^{(\textrm{S})}$ and AMD$^{(\textrm{F})}$ (hatched bands in the middle and bottom pannels, respectively).

\subsection{Statistical Multifragmentation Model}

\begin{figure}
\centering
\includegraphics[width=0.95\hsize]{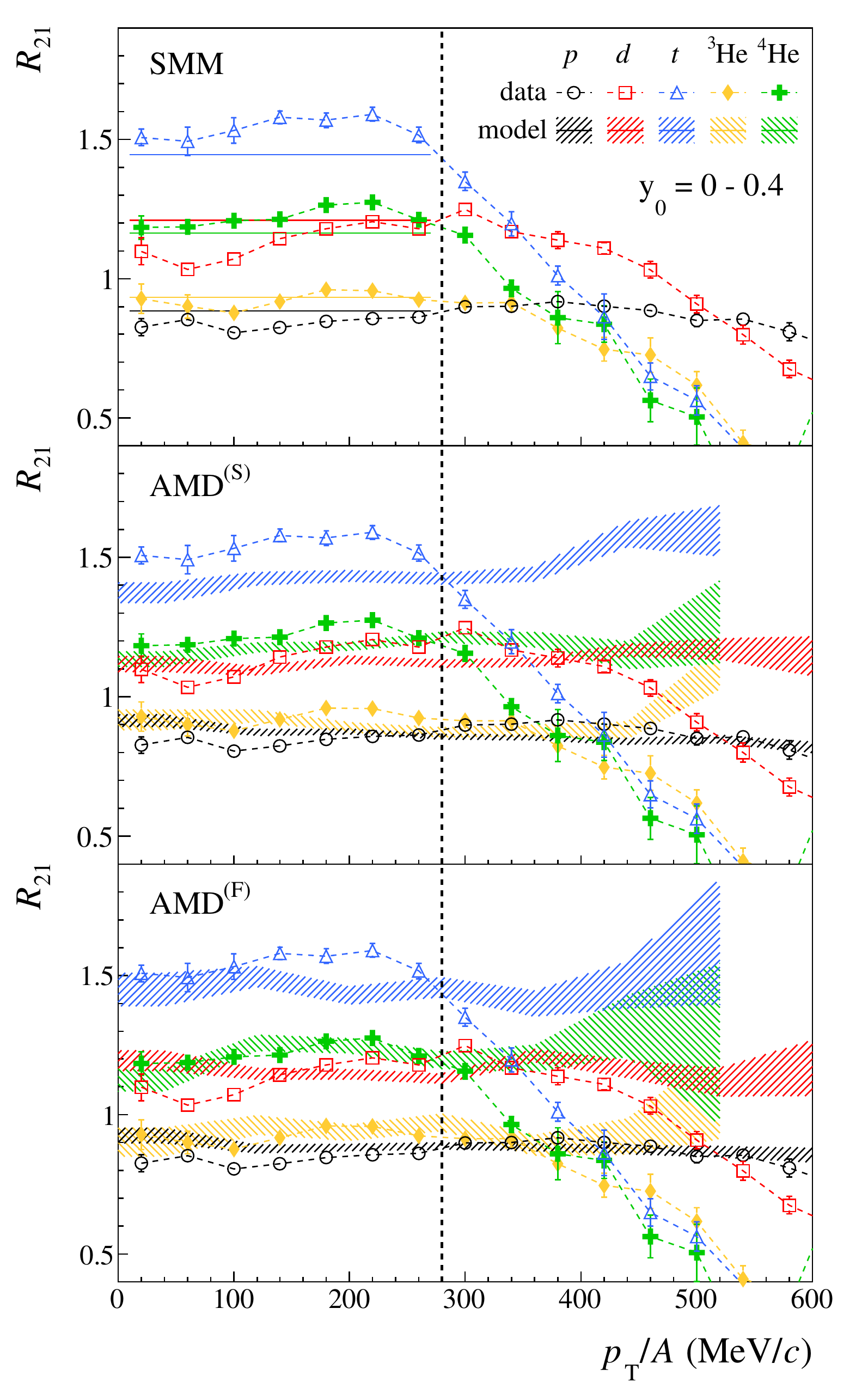}
    \caption{Isotope yield ratios compared with the models: SMM (top), AMD$^{(\textrm{S})}$ (middle) and AMD$^{(\textrm{F})}$ (bottom). The data are the same in all panels. The  break line at $p_T/A$ = 280 MeV/$c$ show  different trend of isoscaling for particles with low (left side of the line) and high (right side of the line) $p_T/A$.}\label{figure_r21}
\end{figure}

In order to check the extent to which the isoscaling properties observed experimentally may be understood in a scenario in which a thermal equilibrated source is formed and undergoes a prompt breakup,
we employ the canonical version of the SMM model described in Refs.~\cite{SouzaSR_NPA_2019_989_69_ISMM,TanWP_PRC_2003_68_034609_ISMM}.
Many different sources contribute to the actual data whereas a single source is employed in the calculation due to the computational effort needed to generate a source distribution.
The model assumes a breakup volume three times larger than that of the source at normal density, and breakup temperature $T=8$ MeV.
The mass and atomic numbers of the decaying source associated with the $^{132}$Sn+$^{124}$Sn system are $N_2=93$ and $Z_2=79$, whereas $N_1=71$ and $Z_1=71$ are used in the case of the $^{108}$Sn+$^{112}$Sn system. 
These values have been selected in order to obtain a good agreement with the measured $R_{21}$ ratios. Different source compositions lead to slightly different $R_{21}$ values and, therefore, those adopted in this work should be seen as average values. The predicted isotope ratios $R_{21}$ are shown as horizontal solid lines in the top panel of Fig.~\ref{figure_r21}.

\subsection{Asymmetrized Molecular Dynamics Model}

In Ref.~\cite{KanekoM_SpiRIT_PLB_2021_822_136681_Z1particles_AMD}, the AMD model~\cite{OnoAkira_PTP_1992_87_1185_AMD,OnoAkira_PPNP_2019_105_139_HIC_dynamics} has been employed to describe the rapidity distributions of $Z$ = 1 particles (top six panels in Fig.~\ref{figure_dmdy}). 
The time evolution of the system is calculated by AMD until $t=300$ fm/$c$. The productions of light charged particles and their properties are almost determined at this primary stage. The employed version of AMD considers the process of cluster formation in the final state of every two-nucleon collision as $N_1+N_2+B_1+B_2\to C_1+C_2$, where each scattered nucleon $N_i$ ($i=1,2$) may form a cluster $C_i$ with a surrounding particle(s) $B_i$. Clusters in the $(0s)^{A_{\text{c}}}$ configuration are considered for $1\le A_{\text{c}}\le 4$. A formed cluster may be broken later, e.g., when a nucleon in the cluster is scattered by some other particle. The important model parameters include the medium effects on two-nucleon collision cross sections $\sigma_{NN}$ and cluster correlations. A set of parameters was chosen in Ref.~\cite{OnoAkira_JPSCP_2020_32_010076_AMD_IMSigmaNN} for a reasonable reproduction of the FOPI data of central Xe + CsI collisions at 250 MeV/nucleon \cite{ReisdorfW_FOPI_NPA_2010_848_366_HICsystematics,AndronicA_PRC_2003_67_034907}, paying attention to the yields of light charged particles and heavier fragments, and a kind of stopping observable of various particle species. (The stopping observable is expressed as $\frac12\langle p_T^2\rangle/\langle p_z^2\rangle$ with transverse and longitudinal momentum components, $p_T$ and $p_z$, in the center-of-mass frame.) This parametrization of Ref. \cite{OnoAkira_JPSCP_2020_32_010076_AMD_IMSigmaNN} is called AMD$^{(\textrm{F})}$ here.

For the present systems of Sn + Sn collisions, the rapidity distributions predicted by AMD$^{(\textrm{F})}$ are shown in the right panels of Fig.~\ref{figure_dmdy}. The calculated distribution looks more transparent than the data and underestimates the yields of deuterons and especially that of tritons in the mid-rapidity region. To reduce this discrepancy in the rapidity distributions in Ref.~\cite{KanekoM_SpiRIT_PLB_2021_822_136681_Z1particles_AMD}, another parametrization by adopting larger in-medium $NN$ cross sections $\sigma_{NN}$ is chosen, so that the rapidity distributions become much narrower as shown in the left panels of Fig.~\ref{figure_dmdy}. In addition, agreement with the triton multiplicity was improved by modifying the phase space so that the bound phase space for the relative coordinate between a two-nucleon pair and another nucleon becomes approximately $(2\pi\hbar)^3$ \cite{KanekoM_SpiRIT_PLB_2021_822_136681_Z1particles_AMD}. We label this parameter set as AMD$^{(\textrm{S})}$. A persistent observed problem is that the $t/{}^{3}\mathrm{He}$ yield ratio is always underestimated, either by AMD$^{(\textrm{F})}$ or AMD$^{(\textrm{S})}$. 

In the present work, we adopt the Skyrme SLy4 effective interaction, which corresponds to a soft symmetry energy with the slope parameter $L=46$ MeV. The calculation of stiff symmetry energy with $L=108$ MeV was also analyzed in Ref.~\cite{KanekoM_SpiRIT_PLB_2021_822_136681_Z1particles_AMD}. However, only the soft symmetry energy case is shown here since the data seems to have a better agreement with the soft symmetry energy case. The interaction with soft symmetry energy overestimates the deuteron binding energy and consequently may overestimate the deuteron yield~\cite{PiantelliS_PRC_2019_99_064616}. However, such an effect of the binding energy or the bound phase space is largely canceled out when the double ratio is taken between two systems~\cite{KanekoM_SpiRIT_PLB_2021_822_136681_Z1particles_AMD}.

In Fig.~\ref{figure_dmdp}, the $p_T/A$ spectra predicted by the AMD$^{(\textrm{S})}$ and AMD$^{(\textrm{F})}$ are depicted by the solid lines (left panels) and dotted lines (right panels) respectively. For proton and deuteron, the overall shapes and absolute yields are well explained with both AMD$^{(\textrm{S})}$ and AMD$^{(\textrm{F})}$. On the other hand, for clusters with $A\ge 3$, the shapes of the $p_T/A$ spectra change significantly when the in-medium cross sections $\sigma_{NN}$ are increased from AMD$^{(\textrm{F})}$ to AMD$^{(\textrm{S})}$, and the spectra extend to high $p_T/A$ in AMD$^{(\textrm{S})}$ compared to the experimental data. The change from AMD$^{(\textrm{F})}$ to AMD$^{(\textrm{S})}$ is understood as a natural consequence of the increased $\sigma_{NN}$ to reduce $\langle p_z^2\rangle$ and increase $\langle p_T^2\rangle$. To reproduce the experimental data of both the rapidity and $p_T/A$ distributions, both $\langle p_z^2\rangle$ and $\langle p_T^2\rangle$ of clusters need to be reduced, in particular for tritons.This is not possible by only changing $\sigma_{NN}$ in AMD. (The enhanced production of $Z=2$ isotope yields from the AMD$^{(\textrm{S})}$ parameterizations in Fig. 2 suggests that adjusting the in-medium cluster correlations may also be needed.)

For the low energy particles with $p_T/A < 280$ MeV/$c$, the behaviour of $R_{21}$ is qualitatively explained by both AMD$^{(\textrm{F})}$ and AMD$^{(\textrm{S})}$. The predicted $R_{21}$ values are shown as bands in the middle and bottom panels of Fig.~\ref{figure_r21}. The widths of the bands represent statistical uncertainties. The predicted $R_{21}$ for all particles except proton show slightly lower values. Similar to SMM, the $R_{21}$ of tritons shows the largest difference in the model comparisons. Furthermore, the $R_{21}$ ratios for tritons from AMD$^{(S)}$ underestimate the data more than the results from AMD$^{(F)}$.

An emergence of isoscaling in the AMD calculation is not trivial, because AMD does not assume any equilibrium state, and the fragments are produced within a rapidly evolving system~\cite{OnoAkira_SpiRIT_PRC_2003_68_051601_isoscaling_AMD}. However, during the dynamical evolution, clusters are repeatedly created and broken by the microscopic processes explained above, and therefore a situation similar to a chemical equilibrium may be realized before particles stop interacting. The isoscaling observed in AMD is consistent with such a picture, at least qualitatively. The precise values of $R_{21}$ may depend on the details of the model ingredients. In the present calculation, all clusters with the $(0s)^{A_{\text{c}}}$ configuration ($1\le A_{\text{c}}\le 4$) are considered, including the dineutron and diproton correlations. When the strength of dinucleon correlation is varied, a test calculation shows that the composition of nucleons and clusters is affected. However, the steep drop of $R_{21}$ of clusters at high $p_T$ down to $R_{21}<1$ in the experimental data seems to be difficult to explain by a minor modification of the present AMD model. Direct information on free neutrons can be an important clue to solve this puzzle. It is available in the calculated results but is not shown in the present paper.

\section{Summary and Conclusion}

In summary, the isoscaling phenomenon of hydrogen and helium isotopes in $^{132}\textrm{Sn}+^{124}\textrm{Sn}$ and $^{108}\textrm{Sn}+^{112}\textrm{Sn}$ reactions at beam energy of 270 MeV/u is presented as a function of $p_T/A$.
Isoscaling phenomenon up to $p_T/A < 280$ MeV/$c$ is found but breaks down for cluster particles with $p_T/A > 280$ MeV/$c$. 
The systems are found to form a thermal equilibrium not throughout the system but locally, which is evident from the increasing trend of H-He isotope ratio temperature with increasing $p_T/A$. The isoscaling can be qualitatively explained by both the SMM and AMD models. When the yield spectra and isoscaling are compared to the predictions from the dynamical model AMD with two different parameter sets, we do not find any preference in increasing the default values of $\sigma_{NN}$ as observed in earlier study of Z=1 particles.
While isoscaling breaks down for $p_T/A > 280$ MeV/$c$ particles in the data,
the isoscaling trend in AMD persist.
Most intriguely, the high-momentum clusters are suppressed in the neutron-rich system compared to the more symmetric system suggesting the non-equilibrium nature of the emission process especially for the high energy particles.

%(comments!)

\section*{Acknowledgement}

The authors would like to thank Prof. Pawel Danielewicz for many fruitful discussions. This work was supported by the U.S. Department of Energy, USA under Grant Nos. DE-SC0021235, DE-NA0003908, DE-FG02-93ER40773, DE-FG02-93ER40773, DE-SC0019209, DE-SC0015266, DE-AC02-05CH11231, U.S. National Science Foundation Grant No. PHY-1565546, the Robert A. Welch Foundation (A-1266 and A-1358), the Japanese MEXT, Japan KAKENHI (Grant-in-Aid for Scientific Research on Innovative Areas) grant No. 24105004, JSPS KAKENHI Grants Nos. JP17K05432, JP19K14709 and JP21K03528,
the National Research Foundation of Korea under grant Nos. 2018R1A5A1025563 and 2013M7A1A1075764, the Polish National Science Center(NCN) under contract Nos. UMO-2013/09/B/ST2/04064, UMO-2013/-10/M/ST2/00624, Computing resources were provided by FRIB, the HOKUSAI-Great Wave system at RIKEN, and the Institute for Cyber-Enabled Research at Michigan State University. 
S.R. Souza acknowledges partial support from CNPq, CAPES, FAPERJ and the use of the supercomputer Lobo Carneiro, where part of the calculations have been carried out. This work has been done as part of the project INCT-F\'\i sica Nuclear e aplica\c c\~oes, projeto No. 464898/2014-5.

\section*{Data Availability Statement} 
This manuscript has no associated data or the data will not be deposited. [Authors' comment: The data can be available on request sent to the corresponding author.]


\begin{thebibliography}{58}
\footnotesize

\bibitem{StockR_PR_1986_135_259}
R. Stock, Particle production in high energy nucleus-nucleus collisions,
Phys. Rep. 135 (1986) 259, \url{https://www.sciencedirect.com/science/article/pii/0370157386901341}.

\bibitem{SengerPeter_PPNP_2004_53_1}
Peter Senger, Particle production in heavy-ion collisions, Progress in Particle and Nuclear Physics 53 (2004) 1, \url{https://doi.org/10.1016/j.ppnp.2004.02.005}.

\bibitem{OnoAkira_PPNP_2019_105_139_HIC_dynamics}
Akira Ono, Dynamics of clusters and fragments in heavy-ion collisions, Progress in Particle and Nuclear Physics 105 (2019) 139, \url{https://doi.org/10.1016/j.ppnp.2018.11.001}.

\bibitem{IkenoNatsumi_PRC_2016_93_044612_AMD_SnSn300}
Natsumi Ikeno, Akira Ono, Yasushi Nara et al., Probing neutron-proton dynamics by pions, Phys. Rev. C 93 (2016) 044612, \url{https://doi.org/10.1103/PhysRevC.93.044612}.

\bibitem{HongJun_PRC_2014_90_024605}
Jun Hong and P. Danielewicz, Subthreshold pion production within a transport description of central Au + Au collisions, Phys. Rev. C 90 (2014) 024605, \url{https://doi.org/10.1103/PhysRevC.90.024605}.

\bibitem{XuHM_PRC_1994_50_1659}
H. M. Xu, W. G. Lynch, and P. Danielewicz, Residue temperatures in intermediate energy nucleus-nucleus collisions, Phys. Rev. C 50 (1994) 1659, \url{https://doi.org/10.1103/PhysRevC.50.1659}.

\bibitem{BorderieB_PPNP_2019_105_82_Phase_Transition_Nuclei}
B. Borderie and J. D. Frankland, Liquid-Gas phase transition in nuclei, Progress in Particle and Nuclear Physics 105 (2019) 82, \url{https://doi.org/10.1016/j.ppnp.2018.12.002}.

\bibitem{WolterHermann_PPNP_2022_125_103962}
Hermann Wolter, Maria Colonna, Dan Cozma, Pawel Danielewicz, Che Ming Ko, Rohit Kumar, Akira Ono, ManYee Betty Tsang, Jun Xu, Ying-Xun Zhang et al., Transport model comparison studies of intermediate-energy heavy-ion collisions, Progress in Particle and Nuclear Physics, 125 (2022) 103962, \href{https://doi.org/10.1016/j.ppnp.2022.103962}{https://doi.org/10.1016/j.ppnp.2022.103962}.

\bibitem{LiBaoAn_PRL_1997_78_1644}
Bao-An Li, C. M. Ko, and Zhongzhou Ren, Equation of State of Asymmetric Nuclear Matter and Collisions of Neutron-Rich Nuclei, Phys. Rev. Lett. 78 (1997) 1644, \url{https://doi.org/10.1103/PhysRevLett.78.1644}.

\bibitem{TanWP_PRC_2001_64_051901}
W. P. Tan, B.-A. Li, R. Donangelo, C. K. Gelbke, M.-J. van Goethem, X. D. Liu, W. G. Lynch, S. Souza, M. B. Tsang, G. Verde, A. Wagner 
and H. S. Xu , Fragment isotope distributions and the isospin dependent equation of state, Phys. Rev. C 64 (2001) 051901, \url{https://doi.org/10.1103/PhysRevC.64.051901}.

\bibitem{JhangG_SpiRIT_PLB_2021_813_136016_pion_ratio}
G. Jhang, J. Estee, J. Barney, G. Cerizza, M. Kaneko, J. W. Lee, W. G. Lynch, T. Isobe, M. Kurata-Nishimura, T. Murakami et al., Symmetry energy investigation with pion production from Sn+Sn systems, Phys. Lett. B 813 (2021) 136016, \url{https://doi.org/10.1016/j.physletb.2020.136016}.

\bibitem{BondorfJP_PR_1995_257_133_SMM}
J. P. Bondorf, A. S. Botvina, A. S. Iljinov, I. N. Mishustin and K. Sneppen, Statistical multifragmentation of nuclei, Phys. Rep. 257 (1995) 133, \url{https://doi.org/10.1016/0370-1573(94)00097-M}.

\bibitem{DasCB_PR_2005_406_1}
C. B. Das, S. Das Gupta, W. G. Lynch and A. Z. Mekjian, M. B. Tsang,
The thermodynamic model for nuclear multifragmentation, Phys. Rep. 406 (2005) 1, \url{https://doi.org/10.1016/j.physrep.2004.10.002}.

\bibitem{BotvinaAS_EPJA_2006_30_121}
A. S. Botvina and I. N. Mishustin, Statistical description of nuclear break-up, Eur. Phys. J. A 30 (2006) 121, \url{https://doi.org/10.1140/epja/i2005-10316-7}.

\bibitem{BotvinaAS_NPA_1987_475_663}
A. S. Botvina, A. S. Iljinov, I. N. Mishustin, J. P. Bondorf, R. Donangelo and K. Sneppen, Statistical simulation of the break-up of highly excited nuclei, Nucl. Phys. A 475 (1987) 663, \url{https://doi.org/10.1016/0375-9474(87)90232-6}.

\bibitem{TanWP_PRC_2003_68_034609_ISMM}
W. P. Tan, S. R. Souza, R. J. Charity, R. Donangelo, W. G. Lynch, and M. B. Tsang, Isospin effects in nuclear multifragmentation, Phys. Rev. C 68 (2003) 034609, \url{https://doi.org/10.1103/PhysRevC.68.034609}.

\bibitem{TsangMB_MSU_PRC_2001_64_054615_isoscaling_SMM_EES}
M. B. Tsang, C. K. Gelbke, X. D. Liu, W. G. Lynch, W. P. Tan, G. Verde, H. S. Xu, W. A. Friedman, R. Donangelo, S. R. Souza et al., Isoscaling in statistical models, Phys. Rev. C 64 (2001) 054615, \url{https://doi.org/10.1103/PhysRevC.64.054615}.

\bibitem{TsangMB_MSU_PRL_2001_86_5023_isoscaling_experiment}
M. B. Tsang, W. A. Friedman, C. K. Gelbke, W. G. Lynch, G. Verde, and H. S. Xu, Isotopic Scaling in Nuclear Reactions, Phys. Rev. Lett. 86 (2001) 5023, \url{https://doi.org/10.1103/PhysRevLett.86.5023}.

\bibitem{XuHS_MSU_PRL_2000_85_716_isoscaling}
H. S. Xu, M. B. Tsang, T. X. Liu, X. D. Liu, W. G. Lynch, W. P. Tan, A. Vander Molen, G. Verde, A. Wagner, H. F. Xi et al., Isospin Fractionation in Nuclear Multifragmentation, Phys. Rev. Lett. 85 (2000) 716, \url{https://doi.org/10.1103/PhysRevLett.85.716}.

\bibitem{SouzaSR_PRC_2009_80_044606_isoscaling_symmetryenergy}
S. R. Souza, M. B. Tsang, B. V. Carlson, R. Donangelo, W. G. Lynch, and A. W. Steiner, Temperature effects in nuclear isoscaling, Phys. Rev. C 80 (2009) 044606, \url{https://doi.org/10.1103/PhysRevC.80.044606}.

\bibitem{RamiF_PRL_2000_84_1120_FOPI_Isospin_tracing}
F. Rami, Y. Leifels, B. de Schauenburg, A. Gobbi, B. Hong, J. P. Alard, A. Andronic, R. Averbeck, V. Barret, Z.Basrak et al. (FOPI Collaboration), Isospin Tracing: A Probe of Nonequilibrium in Central Heavy-Ion Collisions, Phys. Rev. Lett. 84 (2000) 1120, \url{https://doi.org/10.1103/PhysRevLett.84.1120}. 

\bibitem{TsangMB_MSU_PRC_2001_64_041603_isoscaling_condition}
M. B. Tsang, W. A. Friedman, C. K. Gelbke, W. G. Lynch, G. Verde, and H. S. Xu, Conditions for isoscaling in nuclear reactions, Phys. Rev. C 64 (2001) 041603, \url{https://doi.org/10.1103/PhysRevC.64.041603}.

\bibitem{OnoAkira_SpiRIT_PRC_2003_68_051601_isoscaling_AMD}
Akira Ono, P. Danielewicz, W. A. Friedman, W. G. Lynch, and M. B. Tsang, Isospin fractionation and isoscaling in dynamical simulations of nuclear collisions, Phys. Rev. C 68 (2003) 051601, \url{https://doi.org/10.1103/PhysRevC.68.051601}.

\bibitem{DorsoCO_PRC_2006_73_044601}
C. O. Dorso, C. R. Escudero, M. Ison, and J. A. L\'opez,
Dynamical aspects of isoscaling, Phys. Rev. C 73 (2006) 044601, \url{https://link.aps.org/doi/10.1103/PhysRevC.73.044601}.


\bibitem{GeraciE_LNS_NPA_2004_732_173_isoscaling_experiment}
E. Geraci, M. Bruno, M. D'Agostino, E. De Filippo, A. Pagano, G. Vannini, M. Alderighi, A. Anzalone, L. Auditore, V. Baran et al., Isoscaling in central 124Sn+64Ni, 112Sn+58Ni collisions at 35 A MeV, Nucl. Phys. A 732 (2004) 173, \url{https://doi.org/10.1016/j.nuclphysa.2003.11.055}.

\bibitem{TrautmannW_2006_nuclex_0603027_isoscaling_ALADIN_INDRA}
W. Trautmann, A.S. Botvina, J. Brzychczyk, A. Le Fevre, P. Pawlowski, C. Sfienti, and the ALADIN, and INDRA collaborations, Isoscaling and the symmetry energy in spectator fragmentation, International Workshop on Multifragmentation and Related Topics (IWM 2005), \url{https://doi.org/10.48550/arxiv.nucl-ex/0603027}.

\bibitem{FableQ_Arxiv_2022_2202_13850_CaCa_INDRA}
Q. Fable, A. Chbihi, M. Boisjoli, J.D. Frankland, A. Le F\'evre, N. Le Neindre, P. Marini, G. Verde, G. Ademard, L. Bardelli et al., Experimental study of the $^{40,48}$Ca+ $^{40,48}$Ca reactions at 35 MeV/nucleon, Phys. Rev. C 106 (2022) 024605, \url{https://link.aps.org/doi/10.1103/PhysRevC.106.024605}.

\bibitem{WuenschelS_TAMU_PRC_2009_79_061602_isoscaling_experiment}
S. Wuenschel, R. Dienhoffer, G.A. Souliotis, S. Galanopoulos, Z. Kohley, K. Hagel, D. V. Shetty, K. Huseman, L. W. May, S.N. Soisson et al., Isoscaling of fragments with $Z=1\text{\ensuremath{-}}17$ from reconstructed quasiprojectiles, Phys. Rev. C 79 (2009) 061602, \url{https://doi.org/10.1103/PhysRevC.79.061602}.


\bibitem{YoungsM_TAMU_NPA_2017_962_61_isoscaling_experiment}
M. Youngs, A.B. McIntosh, K. Hagel, L. Heilborn, M. Huang, A. Jedele, Z. Kohley, L.W. May, E. McCleskey, A. Zarrella, S.J. Yennello, Observation of different isoscaling behavior between emitted fragments and residues, Nucl. Phys. A 962 (2017) 61, \url{https://doi.org/10.1016/j.nuclphysa.2017.03.009}.

\bibitem{BotvinaAS_JINR_PRC_2002_65_044610_isoscaling_ion_induce}
 A.S. Botvina, O. V. Lozhkin, and W. Trautmann,
 Isoscaling in light-ion induced reactions and its statistical interpretation,
Phys. Rev. C 65 (2002) 044610, \url{https://link.aps.org/doi/10.1103/PhysRevC.65.044610}

\bibitem{LeFevreA_PRL_2005_94_162701_INDRA_Alladin}
A. Le F\`evre, G. Auger, M.L. Begemann-Blaich, N. Bellaize, R. Bittiger, F. Bocage, B. Borderie, R. Bougault, B. Bouriquet, J.L. Charvet, A. Chbihi, {it et al.}, INDRA and ALADIN Collaborations, Isotopic Scaling and the Symmetry Energy in Spectator Fragmentation, Phys. Rev. Lett. 94 (2005) 162701, \url{https://link.aps.org/doi/10.1103/PhysRevLett.94.162701}


\bibitem{ShaneR_SpiRIT_NIMA_2015_784_513_spirittpc}
R. Shane, A. B. McIntosh, T. Isobe, W. G. Lynch, H. Baba, J. Barney, Z. Chajecki, M. Chartier, J. Estee, M. Famiano et al., S$\pi$RIT: A time-projection chamber for symmetry-energy studies, Nucl. Instrum. Methods Phys. Res., Sect. A 784 (2015) 513, \url{https://doi.org/10.1016/j.nima.2015.01.026}.

\bibitem{TangwancharoenS_SpiRIT_NIMA_2017_853_44_TPCGatingGrid}
S. Tangwancharoen, W. G. Lynch, J. Barney, J. Estee, R. Shane, M. B. Tsang, Y. Zhang, T. Isobe, M. Kurata-Nishimura, T. Murakami et al., A gating grid driver for time projection chambers, Nucl. Instrum. Methods Phys. Res., Sect. A 853 (2017) 44, \url{https://doi.org/10.1016/j.nima.2017.02.001}.

\bibitem{BarneyJ_SpiRIT_RSI_2021_92_063302_spirittpc}
J. Barney, J. Estee, W. G. Lynch, T. Isobe, G. Jhang,  M. Kurata-Nishimura,  A. B. McIntosh,  T. Murakami,  R. Shane,  S. Tangwancharoen et al., The S$\pi$RIT time projection chamber, Rev. Scientific Inst. 92 (2021) 063302, \url{https://doi.org/10.1063/5.0041191}.

\bibitem{OtsuH_NIMB_2016_376_175_SAMURAIMagnet}
H. Otsu, S. Koyama, N. Chiga, T. Isobe, T. Kobayashi, Y. Kondo, M. Kurokawa, W. G. Lynch, T. Motobayashi, T. Murakami et al., SAMURAI in its operation phase for RIBF users, Nucl. Instrum. Methods Phys. Res., Sect. B 376 (2016) 175, \url{https://doi.org/10.1016/j.nimb.2016.02.056}.

\bibitem{LaskoP_NIMA_2017_856_92_KATANA}
P. Lasko, M. Adamczyk, J. Brzychczyk, P. Hirnyk, J. \L{}ukasik, P. Paw\l{}owski, K. Pelczar, A. Snoch, A. Sochocka, Z. Sosin et al., KATANA - A charge-sensitive triggering system for the S$\pi$RIT experiment, Nucl. Instrum. Methods Phys. Res., Sect. A 856 (2017) 92, \url{https://doi.org/10.1016/j.nima.2017.03.006}.

\bibitem{KanekoM_NIMA_2022_1039_167010_KyotoArray}
M. Kaneko, T. Murakami, K. Miwa, T. Shiozaki, J. Barney, G. Cerizza, J. Estee, T. Isobe, G. Jhang, M. Kurata-Nishimura et al., Multiplicity trigger detector for the S$\pi$RIT experiment, Nucl. Instrum. Methods Phys. Res., Sect. A 1039 (2022) 167010, \url{https://doi.org/10.1016/j.nima.2022.167010}.


\bibitem{JhangGenie_SpiRIT_JKPS_2016_69_144_SpiRITROOT}
Genie Jhang, Jon Barney, Justin Estee, Tadaaki Isobe, Masanori Kaneko, Mizuki Kurata-Nishimura, Giordano Cerizza, Clementine Santamaria, Jung Woo Lee, Pawe\l{} Lasko et al., {Beam commissioning of the S$\pi$RIT time projection chamber}, J. Korean Phys. Soc. 69 (2016) 144, \url{https://doi.org/10.3938/jkps.69.144}.

\bibitem{LeeJW_SpiRIT_NIMA_2020_965_163840_SpiRITROOT}
J.W. Lee, G. Jhang, G. Cerizza, J. Barney, J. Estee, T. Isobe, M. Kaneko, M. Kurata-Nishimura, W. G. Lynch, T. Murakami et al., Charged particle track reconstruction with S$\pi$RIT Time Projection Chamber, Nucl. Instrum. Methods Phys. Res., Sect. A 965 (2020) 163840, \url{https://doi.org/10.1016/j.nima.2020.163840}.

\bibitem{IsobeT_SpiRIT_NIMA_2018_899_43_GET_electronics}
T. Isobe, G. Jhang, H. Baba, J. Barney, P. Baron, G. Cerizza, J. Estee, M. Kaneko, M. Kurata-Nishimura, J. W. Lee et al., Application of the Generic Electronics for Time Projection Chamber (GET) readout system for heavy Radioactive isotope collision experiments, Nucl. Instrum. Methods Phys. Res., Sect. A 899 (2018) 43, \url{https://doi.org/10.1016/j.nima.2018.05.022}.


\bibitem{TsangCY_SpiRIT_NIMA_2020_959_163477_space_charge}
C. Y. Tsang, J. Estee, R. Wang, J. Barney, G. Jhang, W. G. Lynch, Z. Q. Zhang, G. Cerizza, T. Isobe, M. Kaneko et al., Space charge effects in the S$\pi$RIT time projection chamber, Nucl. Instrum. Methods Phys. Res., Sect. A 959 (2020) 163477, \url{https://doi.org/10.1016/j.nima.2020.163477}.

\bibitem{EsteeJ_SpiRIT_NIMA_2019_944_162509_tpcdynamicrange}
J. Estee, W. G. Lynch, J. Barney, G. Cerizza, G. Jhang, J. W. Lee, R. Wang, T. Isobe, M. Kaneko, M. Kurata-Nishimura et al., Extending the dynamic range of electronics in a Time Projection Chamber, Nucl. Instrum. Methods Phys. Res., Sect. A 944 (2019) 162509, \url{https://doi.org/10.1016/j.nima.2019.162509}.

\bibitem{AndersonM_NIMA_2003_499_659_STARTPC}
M. Anderson, J. Berkovitz, W. Betts, R. Bossingham, F. Bieser, R. Brown, M. Burks, M. Calder\'on de la Barca S\'anchez, D. Cebra, M. Cherney et al., The STAR time projection chamber: a unique tool for studying high multiplicity events at RHIC, Nucl. Instrum. Methods Phys. Res., Sect. A 499 (2003) 659, \url{https://doi.org/10.1016/S0168-9002(02)01964-2}.

\bibitem{EsteeJustinBrian_phdthesis_MichiganStateUniversity_2020}
Justin Brian Estee, CHARGED PION EMISSION FROM NEUTRON-RICH HEAVY ION COLLISIONS FOR STUDIES ON THE SYMMETRY ENERGY, Ph.D thesis, Michigan State University  (2020)

\bibitem{EsteeJ_SpiRIT_PRL_2021_126_162701_pion_ratio}
J. Estee, W. G. Lynch, C. Y. Tsang, J. Barney, G. Jhang, M. B. Tsang, R. Wang, M. Kaneko, J. W. Lee, T. Isobe et al., Probing the Symmetry Energy with the Spectral Pion Ratio, Phys. Rev. Lett. 126 (2021) 162701, \url{https://doi.org/10.1103/PhysRevLett.126.162701}.

\bibitem{KanekoM_SpiRIT_PLB_2021_822_136681_Z1particles_AMD}
M. Kaneko, T. Murakami, T. Isobe, M. Kurata-Nishimura, A. Ono, N. Ikeno, J. Barney, G. Cerizza, J. Estee, G. Jhang et al., Rapidity distributions of Z=1 isotopes and the nuclear symmetry energy from Sn+Sn collisions with radioactive beams at 270 MeV/nucleon, Phys. Lett. B 822 (2021) 136681, \url{https://doi.org/10.1016/j.physletb.2021.136681}.

\bibitem{OnoAkira_PTP_1992_87_1185_AMD}
Akira Ono, Hisashi Horiuchi, Toshiki Maruyama, Akira Ohnishi, Antisymmetrized Version of Molecular Dynamics with Two-Nucleon Collisions and Its Application to Heavy Ion Reactions, Progress of Theoretical Physics 87 (1992) 1185, \url{https://doi.org/10.1143/ptp/87.5.1185}.

\bibitem{CavataC_PRC_1990_42_1760_reduced_impact_parameter}
C. Cavata, M. Demoulins, J. Gosset, M.-C. Lemaire, D. L'H\^ote, J. Poitou, and O. Valette, Determination of the impact parameter in relativistic nucleus-nucleus collisions, Phys. Rev. C 42 (1990) 1760, \url{https://doi.org/10.1103/PhysRevC.42.1760}.

\bibitem{BarneyJonathanElijah_phdthesis_MichiganStateUniversity_2019}
Jonathan Elijah Barney, CHARGED PION EMISSION FROM $^{112}$SN+$^{124}$SN AND $^{124}$SN+$^{112}$SN REACTIONS WITH THE S$\pi$RIT TIME PROJECTION CHAMBER, Ph.D thesis, Michigan State University  (2019)

\bibitem{TsangChunYuen_phdthesis_MichiganStateUniversity_2022}
Chun Yuen Tsang, CONSTRAIN NEUTRON STAR PROPERTIES WITH S$\pi$RIT EXPERIMENT, Ph.D thesis, Michigan State University  (2022)

\bibitem{AlbergoS_INCA_1985_89_1}
S. Albergo, S. Costa, E. Costanzo and A. Rubbino, Temperature and free-nucleon densities of nuclear matter exploding into light clusters in heavy-ion collisions, Nuovo Cimento Soc. Ital. Fis., A 89 (1985) 1, \url{https://doi.org/10.1007/BF02773614}.

\bibitem{KundeGJ_PLB_1998_416_56_HHeTemperature}
G. J. Kunde, S Gaff, C. K. Gelbke, T Glasmacher, M. J. Huang, R. Lemmon, W. G. Lynch, L. Manduci, L. Martin, M. B. Tsang et al., Isospin independence of the H-He double isotope ratio ``thermometer'', Phys. Lett. B 416 (1998) 56, \url{https://doi.org/10.1016/S0370-2693(97)01344-0}.

\bibitem{WangJ_PRC_2005_72_024603}
J. Wang, R. Wada, T. Keutgen, K. Hagel, Y. G. Ma, M. Murray, L. Qin, A. Botvina, S. Kowalski, T. Materna et al., Tracing the evolution of temperature in near Fermi energy heavy ion collisions, Phys. Rev. C 72 (2005) 024603, \url{https://doi.org/10.1103/PhysRevC.72.024603}.

\bibitem{BougaultR_JPG_47_025103}
R Bougault, E Bonnet, B Borderie, A Chbihi,J D Frankland, E Galichet, D Gruyer, M Henri,M La Commara, N Le Neindre et al., Equilibrium constants of hydrogen and helium isotopes at low nuclear densities, J. Phys. G 47 (2020) 025103, \url{https://doi.org/10.1088/1361-6471/ab56ba}.

\bibitem{SouzaSR_NPA_2019_989_69_ISMM}
S. R. Souza, B. V. Carlson, and R. Donangelo, Post breakup dynamical evolution of fragments produced in nuclear multifragmentation, Nucl. Phys. A 989 (2019) 69, \url{https://doi.org/10.1016/j.nuclphysa.2019.05.017}.


\bibitem{OnoAkira_JPSCP_2020_32_010076_AMD_IMSigmaNN}
Akira Ono, Impacts of Cluster Correlations on Heavy-Ion Collision Dynamics, JPS Conf. Proc. 32, 010076 (2020), \url{https://doi.org/10.7566/JPSCP.32.010076}.



\bibitem{ReisdorfW_FOPI_NPA_2010_848_366_HICsystematics}
W. Reisdorf, A. Andronic, R. Averbeck, M. L. Benabderrahmane, O. N. Hartmann, N. Herrmann, K. D. Hildenbrand, T. I. Kang, Y. J. Kim, M. Ki\v{s} et al., Systematics of central heavy ion collisions in the 1A GeV regime, Nucl. Phys. A 848 (2010) 366, \url{https://doi.org/10.1016/j.nuclphysa.2010.09.008}.

\bibitem{AndronicA_PRC_2003_67_034907}
A. Andronic, W. Reisdorf, N. Herrmann, P. Crochet, J.P. Alard, V. Barret, Z. Basrak, N. Bastid, G. Berek, R. \ifmmode \check{C}\else \v{C}\fi{}aplar, {\it et. al.}, FOPI-Collaboration,
Directed flow in Au+Au, Xe+CsI, and Ni+Ni collisions and the nuclear equation of state,
Phys. Rev. C 67 (2003) 034907, \url{https://link.aps.org/doi/10.1103/PhysRevC.67.034907}.

\bibitem{PiantelliS_PRC_2019_99_064616}
S. Piantelli, A. Olmi, P. R. Maurenzig, A. Ono, M. Bini, G. Casini, G. Pasquali, A. Mangiarotti, G. Poggi, A. A. Stefanini et al., Comparison between calculations with the AMD code and experimental data for peripheral collisions of $^{93}\mathrm{Nb}+{\phantom{\rule{0.16em}{0ex}}}^{93}\mathrm{Nb}{,}^{116}\mathrm{Sn}$ at 38 MeV/nucleon, Phys. Rev. C 99 (2019) 064616, \url{https://doi.org/10.1103/PhysRevC.99.064616}.



\end{thebibliography}
\end{document}